\documentclass[12pt]{article}
\usepackage{amssymb}
\usepackage{amsmath}
\usepackage{graphicx}
\def\beq{\begin{equation}}
\def\eeq{\end{equation}}
\usepackage[all,2cell,dvips]{xy}

\def\IR{\relax{\rm I\kern -.18em R}}

\begin{document}
\title{Entanglement of boundary conditions giving rise to spontaneous symmetry breaking in quantum mechanics}
\author{ {\Large A. Restuccia$^{1,2}$, A. Sotomayor$^{3}$ and V. Strauss$^{4}$}}
\maketitle{\centerline{$^1$Departamento de F\'{\i}sica,
Universidad de Antofagasta, Chile}}
\maketitle{\centerline{$^2$Departamento de F\'{\i}sica,
		Universidad Sim\'on Bol\'ivar, Venezuela }}
\maketitle{\centerline {$^3$Departamento de Matem\'{a}ticas,
Universidad de Antofagasta, Chile }}
\maketitle{\centerline {$^{4}$Department of Mathematics,
		State Pedagogical University of Ulyanovsk, Ulyanovsk, Russia}}
\maketitle{\centerline{e-mail: alvaro.restuccia@uantof.cl, adrian.sotomayor@uantof.cl,
vstrauss@mail.ru }}
\begin{abstract}We study spontaneous symmetry breaking in one dimensional quantum mechanical problems in terms of two-point boundary problems which lead to singular potentials containing Dirac delta functions and its derivatives. We search for broken-symmetry bound states. For a particular entanglement of boundary conditions, we show the existence of a ground state, giving rise to a spontaneous symmetry breaking, stable under the phenomenon of decoherence generated from external fluctuations. We discuss the model in the context of the ``chiral" broken-symmetry states of molecules like $NH_3$.

\end{abstract}


Pacs: 02.30. Tb, 03.65 Ge, 11.30. Qc

\section{Introduction}Our main interest in this work is the study of spontaneous symmetry breaking in quantum mechanics and the role of boundary conditions on it. We consider here a one dimensional case. We introduce our proposal by considering a physical relevant quantum mechanical problem in molecular physics.

It is well known that the stable  quantum state of many molecules in nature, when the configuration of atoms is symmetric under parity, breaks this symmetry. The state is left handed or right handed. The physical argument explaining this symmetry breaking is also well established. Let us consider the molecule $NH_3$ as an example, and consider $x$ to be the axis perpendicular to the plane determined by the three nucleus of the $H$ atoms. The distribution of the electron pairs making the chemical bond between the atoms is:
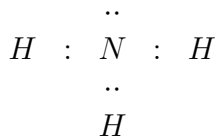
\begin{eqnarray*}\begin{array}{ccccc}&&\cdot \cdot&&\\H & : & N & : & H\\&&\cdot
		\cdot&&\\&&H&& \end{array}\end{eqnarray*}

The wave function of the $N$ ion along the $x$ axis, following the usual argument, is determined by solving the eigenvalue problem for a Schr\"odinger operator with a symmetric local potential. The potential has two symmetric minima representing the attractive potentials of the electron pairs and a repulsive barrier between them representing the action of the three positive $H$ nucleuses.

The Hamiltonian is symmetric, and the ground state for this quantum mechanical problem is non-degenerate. Consequently, the ground state has to be symmetric (even) or antisymmetric (odd). In fact, if $u(x)$ is a solution of the eigenvalue problem so is $u(-x)$ with the same eigenvalue and since there is no degeneracy then $u(x)=au(-x)$ with $a$ equal to 1 or -1. However, this ground state does not corresponds to what is experimentally known from the molecule of $NH_3$, the state of the molecule corresponds to a broken symmetry state with the wave function concentrated on one  side or the other of the repulsive barrier. The argument, see for example \cite{Weinberg}, justifying that the molecule is in a state which is not an eigenstate, but that it naturally arises by perturbation of an infinitely high and thick repulsive barrier, is based on the phenomenon of decoherence (for a review see \cite{Zurek}). External perturbations change the phase of a symmetric or antisymmetric state giving rise to an unstable incoherence mixture of both. On the other side a state concentrated only on one side of the repulsive barrier will be stable under external phase fluctuations. Although the broken-symmetry state is insensitive to external perturbations it is not stable under time evolution, because it is not an eigenstate. In fact, a left handed state evolves, according to the Schr\"odinger equation (ignoring the decoherence phenomena), from the left handed state to a right handed one and viceversa, but it may decay during such evolution to an incoherence mixture of left and right handed states when external perturbations are included in the process. However, if the repulsive barrier is high enough this decaying process may take so long to be unobservable. This argument explains why the right or left handed states, although they are not eigenstates of the Hamiltonian, are the ones existing in nature.

We will analyse in this work the one dimensional quantum problem above from a different poin of view. We will introduce a quantum interaction which is not described by a local potential acting on the wave function as a product. It acts as a linear operator on the wave function, but it cannot be described as a product. Its image will be a linear combination of two derivatives of the Dirac delta at symmetric points separated $h$ from the origin with coefficients depending on the values of the wave function at $h$ and at $-h$ in a way that the Hamiltonian will be symmetric under parity transformation, self adjoint and non-local in the sense that the coefficients of the derivative of the Dirac delta at $h$ depend on the wave function at $-h$ and viceversa. That is, an entanglement of boundary conditions. The mathematical procedure we will follow is a method of self-adjoint extensions of a symmetrical operator. The benefit of introducing this exotic potential is that the ground state will be a broken-symmetry state stable under external fluctuations, and since it is an exact eigenfunction of the Hamiltonian it is also stable under time evolution, in distinction to what occurs in the previous argument. Moreover, this occurs even when the ``potential" has a finite coupling constant in distinction to what happens in local interactions where one needs and infinitely high thick barrier to achieve a broken symmetry eigenstate.

The use of the Dirac delta distribution to construct approximate models describing physical interesting problems has been performed in the past with remarkable success  \cite{Burrau,Yang,Kronig,Frost1,Frost2,Ahlrichs,Certain,Albeverio1,Gesztesy}. One of them is an approximate model of an hydrogen molecule ion, $H_2^+$. The ion is commonly formed in molecular clouds in space. The first successful quantum mechanical treatment of $H_2^+$ was published by O. Burrau in 1927 \cite{Burrau}. The electron of the molecule moves in an attractive potential generated by the two protons. The quantum analysis of the electron arises from the solution of the Schr\"odinger equation, where the potential can be approximated by two delta functions with negative coefficients describing the attractive interaction.
Another interesting application arises from an approximation to the Kronig-Penney model \cite{Kronig}, which describes some basic quantum effects in the conduction of electrical charges in metals. The potential is expressed as an infinite sequence of Dirac deltas with a finite separation between them, a combination of Dirac deltas. They have been also used in toy models of quantum wires or nanowires.
In these applications the potential, expressed in terms of the Dirac delta distribution at some points, is local in the sense that the coefficients of the delta are the values of the wave function at those points. Although our proposal is in terms of derivative of the Dirac delta distribution the new point is on the entaglement of boundary conditions.

In section 2 we present preliminary results on the study of singular potentials in terms of delta functions. In particular, we discuss a Hamiltonian whose ground state is degenerate, with two eigenfunctions, however there is no symmetry breaking in this case. Next, in section 3 we  go to the main contribution of the paper by introducing local and non-local potentials and discuss their quantum properties. In particular, the spontaneous symmetry breaking. Finally, we  give our conclusions in section 4.

\section{Preliminaries}Let us recall a simple case of one-point interaction.
For this we consider the differential
operator $D=-d^{2}\cdot /dx^2$ acting on the real line $\mathbb{R}$ with
the domain  \[{\mathcal D}(D)=\{ y(x)|\;
y(x),y^{\prime}(x),y^{\prime\prime}(x)\in L^{2}({\mathbb R}), \,
y(0)=y^{\prime}(0)=0\}.\]
Then the adjoint operator $D^*$ has
the domain \[{\mathcal D}(D^*)=\{ y(x)|\;
y(x),y^{\prime}(x),y^{\prime\prime}(x)|_{{\mathbb R}_{+}}\in L^{2}({\mathbb
	R}_{+}), \, y(x),y^{\prime}(x),y^{\prime\prime}(x)|_{{\mathbb{R}}_{-}}\in
L^{2}({\mathbb R}_{-})\},\]
where $ {\mathbb R}_{+}=\{ x|\; x\geq
0\}$ and  ${\mathbb R}_{-}=\{ x|\; x\leq 0\}. $

Since $D\subset D^{*}$, for
every self-adjoint extension $\widetilde{D}$ of $D$ we have
$D\subset \widetilde{D}\subset D^{*}$ and an extension of $D$ can
be reduced to a restriction of $D^{*}$. For $D^{*}$ we have
$(D^{*}y,z)=-y^{\prime}(-0)\bar{z}(-0)
+y^{\prime}(+0)\bar{z}(+0)+y(-0)\bar{z}^{\prime}(-0)
-y(+0)\bar{z}^{\prime}(+0)+(y,D^{*}z)$.

We then obtain a self-adjoint restriction given by \beq
y^{\prime}(-0)\bar{z}(-0)
-y^{\prime}(+0)\bar{z}(+0)-y(-0)\bar{z}^{\prime}(-0)
+y(+0)\bar{z}^{\prime}(+0)=0.\label{1}\eeq
The space of boundary values in our
case is four-dimensional, so for a self-adjoint restriction we
need two linear homogenous conditions. One of these conditions
have the form (the same for $y$ and $z$) $y(-0)=y(+0)$ and
$\frac{y^{\prime}(+0)-y^{\prime}(0)}{y(0)}=c =const$. In this case
the first generalized derivative of $y(t)$ has a jump and, so, the
second one has a generalized summand with the delta-function. Thus
the corresponding extension $\widetilde{D}$ can be naturally
presented in the form $\widetilde{D}y(x)=-d^{2}y(x)/dx^{2}+c\cdot
y(0)\delta (x)=-d^{2}y(x)/dx^{2}+c\cdot y(x)\delta (x)$. If $c<0$,
then $\widetilde{A}$ has the negative eigenvalue $\lambda =-c^{2}$
that corresponds to the eigenfunction $y(t)=e^{-c|x|}$. These
facts are well known and can be find in the book of S. Albeverio, F. Gesztesy, R. Hoegh-Krohn, H. Holden \cite{Albeverio2}.  Here they also study singular potentials with derivatives of the shifted delta function in finitely many points, but they only consider local boundary conditions.

In the present paper we  study some generalization of the
described above scheme for two  boundary problems
for points $-h,h$  and a
behaviour of the corresponding extension if $h\to 0$. We will show
that this behaviour involves the derivatives of the delta-function.

Note that even for the one-point problem there are some
self-adjoint extensions with one or two negative eigenvalues, they
naturally involve not only the delta-function but its first
derivative. It is easy to check that for the extension   given by the boundary
conditions ($\alpha >0,\; \beta >0$)
\beq
\begin{array}{ll}y(+0)=\displaystyle\frac{1}{2}\left\{-\left(\frac{1}{\alpha}+
\frac{1}{\beta}\right) y^{\prime}(+0)+
\left(\frac{1}{\alpha}-\frac{1}{\beta}\right) y^{\prime}(-0)\right\},\\
y(-0)=\displaystyle\frac{1}{2}\left\{\left(-\frac{1}{\alpha}+\frac{1}{\beta}\right)
y^{\prime}(+0)+ \left(\frac{1}{\alpha}+\frac{1}{\beta}\right)
y^{\prime}(-0)\right\},\end{array}\label{2}\eeq
where
$\int_{-\infty}^{-0}(|y(x)|^{2}+|y(x)^{\prime}|^{2}+|y(x)^{\prime\prime}|^{2})dx
+
\int_{+0}^{+\infty}(|y(x)|^{2}+|y(x)^{\prime}|^{2}+|y(x)^{\prime\prime}|^{2})dx
<\infty$,\ conditions (\ref{1}) are fulfilled, therefore it is self-adjoint. A direct verification shows that
the eigenvalues for this extension are $-\alpha^2$ and $-\beta^2$,
the corresponding eigenfunctions are $e^{-\alpha |x|}$ and
$sign(x)e^{-\beta |x|}$ respectively. The extension $\widetilde{D}$, that is the self-adjoint Hamiltonian,
has the representation \[\widetilde{D}y(x)=-y^{\prime\prime}(x)-\frac{1}{\beta }\cdot\delta^{\prime}(x)
\left( y^{\prime}(-0)+y^{\prime}(+0)\right) -{\alpha }
\cdot\delta(x)\left( y(-0)+y(+0)\right).\]

An interesting case occurs when $\alpha =\beta$ because
the latter means that the unique negative eigenvalue
$-\alpha^2$ has two non symmetric eigenfunctions

\begin{eqnarray*} y_1(t)=\left\{\begin{array}{lll} e^{\alpha x}\hspace{2mm}\mathrm{\:if\:}\hspace{2mm}
		x<0
		\\ \\ 0 \hspace{2mm}\mathrm{\:if\:}\hspace{2mm}
		x>0
	\end{array}\right.\end{eqnarray*} and

\begin{eqnarray*} y_2(x)=\left\{\begin{array}{lll} 0\hspace{2mm}\mathrm{\:if\:}\hspace{2mm}
		x<0
		\\ \\  e^{-\alpha x}\hspace{2mm}\mathrm{\:if\:}\hspace{2mm}
		x>0
	\end{array}\right.\end{eqnarray*}

However, the case under consideration cannot be interpreted as a model giving rise to an spontaneous symmetry breaking. In fact, this is the case of a non transitable barrier because the conditions (\ref{2}) imply
\[\alpha\cdot y(-0)=y^{\prime}(-0)\mbox{ and }\alpha\cdot y(+0)=-y^{\prime}(+0)\ .\]
The latter means that waves on $\mathbb{R}_-$ and $\mathbb{R}_+$ are independent.

\section{Main construction}
\subsection{A general idea}
Let  the differential operator $D=-d^{2}\cdot /dx^2$ acting on the real
line $\mathbb R$ with domain \[{\mathcal D}(D)= \{ y(x)|\;
y(x),y^{\prime}(x),y^{\prime\prime}(x)\in L^{2}({\mathbb R}), \,
y(-h)=y^{\prime}(-h)=y(h)=y^{\prime}(h)=0\}.\]

Then for the adjoint operator $D^{*}$ we have
\begin{eqnarray*}&&(D^{*}y,z)-(y,D^{*}z)=-y^{\prime}(-h-0)\bar{z}(-h-0)
+y^{\prime}(-h+0)\bar{z}(-h+0)\\&& -y^{\prime}(h-0)\bar{z}(h-0)
+y^{\prime}(h+0)\bar{z}(h+0)+y(-h-0)\bar{z}^{\prime}(-h-0)
-y(-h+0)\bar{z}^{\prime}(-h+0)\\&& +y(h-0)\bar{z}^{\prime}(h-0)
-y(h+0)\bar{z}^{\prime}(h+0).\end{eqnarray*}

Thus, a restriction of $D^{*}$ would be selfadjoint for any minimally restrictive boundary conditions such that
\beq\begin{array}{lll}
-y^{\prime}(-h-0)\bar{z}(-h-0)
+y^{\prime}(-h+0)\bar{z}(-h+0)  \\-y^{\prime}(h-0)\bar{z}(h-0)
+y^{\prime}(h+0)\bar{z}(h+0) \\+y(-h-0)\bar{z}^{\prime}(-h-0)
-y(-h+0)\bar{z}^{\prime}(-h+0)\\ +y(h-0)\bar{z}^{\prime}(h-0)
-y(h+0)\bar{z}^{\prime}(h+0)  =0.\end{array}\label{3}\eeq

Seeking self-adjoint extensions of $D$ we can, for instance, assume that
\beq y(-h-0)=y(-h+0),\quad
y(h-0)=y(h+0)\label{4}\eeq
and the same for $z(x)$, i.d. the continuity of $y$ and $z$. Then
the conditions of selfadjointness for extensions of $D$ convert
to the equality \begin{eqnarray*}&&
(y^{\prime}(-h+0)-y^{\prime}(-h-0))\bar{z}(-h)
+(y^{\prime}(h+0)-y^{\prime}(h-0))\bar{z}(h)  \\
&&-y(-h)(\bar{z}^{\prime}(-h+0)
-\bar{z}^{\prime}(-h-0))-y(h)(\bar{z}^{\prime}(h+0)
-\bar{z}^{\prime}(h-0))  =0\ .\end{eqnarray*}
This case does not bring qualitatively new effects in comparison to the one presented in the section Preliminaries.

\subsection{The entanglement of boundary conditions}
Seeking self-adjoint extensions of $D$ let us suppose that
\beq y^{\prime}(-h-0)=y^{\prime}(-h+0),\quad
y^{\prime}(h-0)=y^{\prime}(h+0)\label{5}\eeq and the
same for $z(x)$. Then the conditions of self-adjointness for
extensions of $D$ convert to
\begin{eqnarray*}&&y^{\prime}(-h)\left(-\bar{z}(-h-0) +\bar{z}(-h+0)\right)
+y^{\prime}(h)\left(-\bar{z}(h-0) +\bar{z}(h+0)\right) + \\ &&\left(
y(-h-0) -y(-h+0)\right)\bar{z}^{\prime}(-h) +\left( y(h-0)
-y(h+0)\right)\bar{z}^{\prime}(h)=0.\end{eqnarray*}Let us seek
self-adjoint extensions such that

\beq\begin{pmatrix}(y(-h+0)-y(-h-0))\cr
(y(h+0)-y(h-0))\end{pmatrix}=\begin{pmatrix}b_{11}& b_{12}\cr b_{21}&
b_{22}\end{pmatrix}\begin{pmatrix}y^{\prime}(-h)\cr
y^{\prime}(h)\end{pmatrix}.\label{6}\eeq

It is easy to check that the
corresponding extension will be self-adjoint if and only if the
matrix
\beq B=\begin{pmatrix}b_{11}& b_{12}\cr b_{21}& b_{22}\end{pmatrix}\label{7}\eeq
is symmetric.

Let us consider a function $y(x)$ under conditions (\ref{6})  as a generalized function (distribution). Then
\begin{equation*}
y^{\prime\prime}(x)=y^{\prime\prime}_{cl}(x)+(b_{11}y^{\prime}(-h)+
b_{12}y^{\prime}(h))\delta^{\prime}(x+h)
+(b_{21}y^{\prime}(-h)+b_{22}y^{\prime}(h))\delta^{\prime}(x-h)\ ,
\end{equation*}
where
\begin{equation*} f_{cl}^{\prime\prime}(x)=\begin{cases}f^{\prime\prime}(x),& \text{ if } f^{\prime\prime}(x)
\text{ exists in the classical sense}\\ 0,& \text{ in the opposite case.} \end{cases}\end{equation*}

Thus, the corresponding extension $\widetilde{D}_h$ of the operator $D$ can be re-written as
\beq\begin{array}{ll}
\widetilde{D}_hy(x)= & -y^{\prime\prime}(x)+(b_{11}y^{\prime}(-h)+b_{12}y^{\prime}(h))
\delta^{\prime}(x+h)
\\ & +(b_{21}y^{\prime}(-h)+b_{22}y^{\prime}(h))\delta^{\prime}(x-h)\ .\\
\end{array}\label{8}\eeq

This is the image of the Hamiltonian, for a suitable matrix $B$ which we will determine below, of our proposal acting on a wave function $y(x)$ for the quantum mechanical problem discussed in the introduction based on the molecular structure of $NH_3$. When $b_{11}=b_{22}$ and
$b_{12}=b_{21}$ it is invariant under parity transformations, in fact, if $u(x)$ is an eigenfunction so is $u(-x)$ with the same eigenvalue.

Let us construct a matrix $B$ such that for every positive $h$ the
function
\begin{equation} f_{h}(x)=\begin{cases}e^{\alpha x},& x\leq -h\\
-{\displaystyle\frac{e^{-\alpha h}(e^{-\alpha x}+e^{\alpha
			x})}{e^{\alpha h}- e^{-\alpha h}}},& x\in (-h,h)\\ e^{-\alpha
	x},& x\geq h\end{cases}\label{9}\end{equation} would be an
eigenfunction of the operator $\widetilde{D}_h$. It is clear that
the corresponding eigenvalue is $\lambda =-\alpha^2$. Note that
$\int_{-h}^{h}f_{h}(x)dx=-\frac{2e^{-\alpha h}}{\alpha}$, so in  the sense of distributions
$\lim_{h\to +0}f_{h}(x)=f_{0}(x)-\frac{2}{\alpha}\delta (x)$,
where $f_{0}(x)=e^{-\alpha^{2}|x|}$, therefore in this case the limit
generates a new boundary problem, that directly involves $\delta (x)$.

Let us show that there is an extension $\widetilde{D}_h$ that corresponds
to $f_{h}(x)$.

According to (\ref{6}) we have
$$-\frac{2e^{-\alpha h}}{1-e^{-2\alpha h}}=\alpha \cdot
e^{-\alpha h}\cdot (b_{11}-b_{12}),$$
$$\frac{2e^{-\alpha h}}{1-e^{-2\alpha h}}=\alpha \cdot
e^{-\alpha h}\cdot (b_{21}-b_{22}).$$

The last system implies, using that $B$ is symmetric, $b_{11}=b_{22}$ and determines $b_{21}-b_{12}$.

Since $B$ is not completely determined we can further impose that
$\widetilde{D}_h$ has another eigenvalue $\mu =-\beta^2$. If an
eigenfunction $g_{h}(x)$ corresponds to $\mu$, then it  must be
orthogonal to $f_{h}(x)$. This condition; in particular, would be fulfilled if $g_{h}(x)$ is odd. We consider
\begin{equation} g_{h}(x)=\begin{cases}e^{\beta x},& x\le -h\\
{\displaystyle\frac{e^{-\beta h}(-e^{-\beta x}+e^{\beta
			x})}{e^{\beta h}+ e^{-\beta h}}},& x\in (-h,h)\\ -e^{-\beta
	x},& x\geq h.\end{cases}\label{10}\end{equation}
According to (\ref{6}) we have

$$-\frac{2e^{-\beta h}}{1+e^{-2\beta h}}=\beta \cdot
e^{-\beta h}\cdot (b_{11}+b_{12}),$$
$$-\frac{2e^{-\beta h}}{1+e^{-2\beta h}}=\beta \cdot
e^{-\beta h}\cdot (b_{21}+b_{22}).$$
So
\begin{equation}\begin{array}{lcr}
b_{22}=b_{11}&=&\displaystyle -\left(\frac{1}{\alpha (1-e^{-2\alpha h})}+
\frac{1}{\beta (1+e^{-2\beta h})}\right),\\
b_{21}=b_{12}&=&\displaystyle\left(\frac{1}{\alpha (1-e^{-2\alpha h})}-
\frac{1}{\beta (1+e^{-2\beta h})}\right).\end{array}\label{11}\end{equation}

\par This interaction is local if and only if $b_{12}=b_{21}=0$, that means
$$  \frac{1}{\alpha (1-e^{-2\alpha
		h})}= \frac{1}{\beta (1+e^{-2\beta h})}.$$
The latter implies $\beta < \beta\cdot (1+e^{-2\beta h})=\alpha\cdot (1-e^{-2\alpha h}) <\alpha $, so in the case  of a local interaction the ground
state is given by (\ref{9}).

The most interesting case occurs when we consider $\alpha=\beta$; then
\begin{equation}
b_{11}=\displaystyle \frac{-2}{\alpha (1-e^{-4\alpha h})}
,\
b_{12}=\displaystyle\frac{2e^{-2\alpha h}}{\alpha (1-e^{-4\alpha h})}.\label{12}\end{equation}
Thus,  $b_{12}>0$ and we have an entanglement of boundary conditions.  The ground state is a linear combination of  (\ref{9}) and
 (\ref{10}) and can be asymmetric. In fact, the ground state is degenerate, (\ref{9}) and
 (\ref{10}) are both eigenfunctions as well as any linear combination of them. In particular, $f_h+g_h$ is an eigenstate with the wave function concentrated on $x\leq h$ and $f_h-g_h$  is an eigenstate with the wave function concentrate on $x\ge -h$. In addition, when we consider the phenomenon of decoherence the only stable eigenstate under external fluctuations corresponds to $f_h+g_h$ or $f_h-g_h$, the left handed or right handed ground states. In distinction to what occurs in the argument presented in the introduction, the stable state is an exact eigenfunction and hence it is also stable under the time evolution. Consequently, the entanglement of boundary conditions gives rise to spontaneous symmetry breaking of the parity symmetry of the Hamiltonian (\ref{8}).

  Note that $b_{11},b_{12}\to 0$ and $\frac{b_{12}}{b_{11}}\to 0$ if $\alpha\to\infty$, so for relatively big $\alpha$
the violation of locality in (\ref{6}) is relatively weak. In the opposite case, if $\alpha\to 0$, then
$b_{11},b_{12}\to \infty$ and $\frac{b_{12}}{b_{11}}\to -1$, so in this case the entanglement between the points $-h$ and $h$
is quite strong. The same effect takes place if $\alpha$ is a constant but $h\to 0$, that is natural.
\par
Returning to the general case of the relation between $\alpha$ and $\beta$ let us note that the expression (\ref{8})
can be reexpressed as
\begin{equation*}\label{d1}\begin{array}{lr}
\widetilde{D}_hy(x)= & -y^{\prime\prime}(x)+
\frac{(\delta^{\prime}(x-h)-\delta^{\prime}(x+h))(y^{\prime}(-h)-y^{\prime}(h))}{\alpha (1-e^{-2\alpha h})}\\ & -
\frac{(\delta^{\prime}(x+h)+\delta^{\prime}(x-h))(y^{\prime}(-h)+y^{\prime}(h))}{\beta (1+e^{-2\beta h})} .\end{array}
\end{equation*}
Strictly speaking in the latter expression we cannot pass to the limit $h\to 0$ because the domain of $\widetilde{D}_h$ depends on $h$
and, moreover, the eigenfunction $f_h(x)$ does not converge to any function in $L^2(\mathbb{R})$.

\section{Conclusions}

In this work, we introduced a new quantum interaction of the type of derivatives of two shifted Dirac delta distributions with coefficients which ensure that the Hamiltonian operator is self adjoint. It results from the analysis of a method of self adjoint extensions of symmetric (with respect to the internal product) operators. Among the admisible boundary conditions there is a class of them which renders the Hamiltonian invariant under parity transformations. Within this class there are boundary conditions for which the interaction is local. In this case the discrete spectrum consists of an even or symmetric (under parity) state and an odd or antisymmetric one, the ground state corresponding to the symmetric state. There is also a class of boundary conditions which determines a non-local interaction. We called it an entanglement of boundary conditions. In this case the ground state is degenerate and there are eigenstates with the wave function concentrated on one side of the interarction zone, bein zero on its complement. These are the left or right handed states. Any linear combination of them is also eigenstate with the same eigenvalue, however when external perturbations on the wave function are considered, which inevitable occur, the only stable states are just the left or right handed ones. The other eigenstates, because of the decoherence phenomenon, rapidly become an incoherence mixture of even and odd states. The entanglement of boundary conditions gives rise then to a spontaneous symmetry breaking. It is interesting that this effect is obtained for a finite coupling constant on the derivative of the Dirac delta distribution, in distinction to the case of a local interaction where the only possibility to have spontaneous symmetry breaking occurs for an infinitely high and thick barrier. If we represent it in terms of a local Dirac delta interaction, the coupling of it must neccesarily approach infinity.

We applied the non-local interaction to give a qualitative description of the molecular structure of $NH_3$. We compared our argument based  on exact energy eigenstates to the well established one for which the left and right handed states are only nearly energy eigenstates.

$\bigskip$

\textbf{Acknowledgments}

A. R. and A. S.  are partially supported by Project Fondecyt 1161192, Chile.

\end{document}